\documentclass[iop]{emulateapj}
\usepackage[usenames,dvipsnames]{color}
\usepackage[draft]{changes}

\newcommand{\mem}[1]{\ensuremath{\mathrm{ #1}}}
\newcommand{\n}{\ensuremath{\mem{n}}}
\newcommand{\hefo}{\ensuremath{^{4}\mem{He}}}
\newcommand{\czw}{\ensuremath{^{12}\mem{C}}}
\newcommand{\cdr}{\ensuremath{^{13}\mem{C}}}

\newcommand{\ndr}{\ensuremath{^{13}\mem{N}}}

\newcommand{\ose}{\ensuremath{^{16}\mem{O}}}

\newcommand{\ipr}{\mbox{i-process}}
\newcommand{\snia}{\mbox{SN\,Ia}}
\newcommand{\sneia}{\mbox{SNe\,Ia}}
\newcommand{\mppnp}{\textsc{mppnp}}
\newcommand{\mesa}{\textsc{mesa}}
\newcommand{\iprn}{\mbox{i\,process}}
\newcommand{\natlog}[2]{\ensuremath{#1 \cdot 10^{#2}}} 
\newcommand{\sect}[1]{\S\,\ref{#1}}
\newcommand{\be}{\begin{displaymath}}
\newcommand{\ee}{\end{displaymath}}
\newcommand{\bea}{\begin{eqnarray}}
\newcommand{\eea}{\end{eqnarray}}
\newcommand\msol{M_{\odot}}

\newcommand{\fig}[1]{Fig.\,\ref{#1}}
\newcommand{\tab}[1]{Table\,\ref{#1}}

\usepackage{natbib}

\shortauthors{Denissenkov, Herwig, Battino et al.}
\shorttitle{\iprn\ on rapidly accreting white dwarfs}

\begin{document}

\title{ i-process nucleosynthesis and Mass Retention Efficiency 
  in He-shell
  flash evolution of Rapidly Accreting White Dwarfs}

\author{Pavel A. Denissenkov\altaffilmark{1,2,3}, Falk Herwig\altaffilmark{1,2,3},
  Umberto Battino\altaffilmark{3,4}, Christian Ritter\altaffilmark{1,2,3},
  Marco Pignatari\altaffilmark{3,5,6}, Samuel Jones\altaffilmark{3,7} and Bill Paxton\altaffilmark{8}}
\altaffiltext{1}{Department of Physics \& Astronomy, University of Victoria,
  P.O.~Box 1700, STN CSC, Victoria, B.C., V8W~2Y2, Canada}
\altaffiltext{2}{Joint Institute for Nuclear Astrophysics, Center for the Evolution of the Elements, Michigan
  State University, 640 South Shaw Lane, East Lansing, MI 48824, USA}
\altaffiltext{3}{NuGrid collaboration, \url{http://www.nugridstars.org}}
\altaffiltext{4}{Department of Physics, University of Basel, Klingelbergstrasse 82, CH-4056 Basel, Switzerland}
\altaffiltext{5}{E.A. Milne Centre for Astrophysics, Department of Physics \& Mathematics, University of Hull, HU6 7RX, United Kingdom}
\altaffiltext{6}{Konkoly Observatory, Research Centre for Astronomy and Earth Sciences, Hungarian Academy of Sciences, Budapest, Konkoly Thege Mikl\'{o}s \'{u}t 15-17, 1121 Budapest, Hungary}
\altaffiltext{7}{Heidelberg Institute for Theoretical Studies, Schloss-Wolfsbrunnenweg 35, 69118 Heidelberg, Germany }
\altaffiltext{8}{Kavli Institute for Theoretical Physics and Department of Physics, Kohn Hall, University of California,
  Santa Barbara, CA 93106, USA}
 
\begin{abstract}
Based on stellar evolution simulations, we demonstrate that rapidly
accreting white dwarfs  in close binary systems are an astrophysical site for the intermediate
neutron-capture process. During recurrent and very strong He-shell flashes in the stable
H-burning accretion regime H-rich
material enters the He-shell flash convection zone. $\czw(p,\gamma)\ndr$ reactions release enough energy 
to potentially impact convection, and \iprn\ is activated through the $\cdr(\alpha,\n)^{16}$O reaction.  The H-ingestion flash may not cause a split of the convection
zone as it was seen in simulations of He-shell flashes in post-AGB and low-Z AGB stars.
We estimate that
for the production of first-peak heavy elements this site can be of similar importance for
galactic chemical evolution as the s-process production by low-mass
AGB stars.
The He-shell flashes result in the
expansion and, ultimately, ejection of the accreted and then i-process enriched material, via
super-Eddington luminosity winds or Roche-lobe overflow. The white
dwarf models do not retain any significant amount of the
accreted mass, with a He retention efficiency of $\la 10\%$ depending
on mass and convective boundary mixing assumptions. This makes the evolutionary
path of such systems to supernova Ia explosion highly
unlikely. 

\end{abstract} 


\section{Introduction}
\label{s.introduction}
Trans-iron elements are produced through n-capture nucleosynthesis,
such as the slow (s, with a neutron density $N_\mathrm{n}\la 10^{11}\,
\mathrm{cm}^{-3}$) and the rapid (r, $N_\mathrm{n}\ga 10^{20}\,
\mathrm{cm}^{-3}$) processes
\citep[e.g.,][]{kaeppeler:11,thielemann:11}.
\cite{cowan:77} proposed that an \iprn\ with $N_\mathrm{n}\sim
10^{15}\, \mathrm{cm}^{-3}$ intermediate between s and r processes
might be triggered when H is mixed into a convective He-burning shell.
Neutrons for the \iprn\ are released in the reaction
$^{13}$C($\alpha$,n)$^{16}$O. The high n density in the \iprn\ is
achieved at typical He-burning temperatures in the range $2$ to
$\natlog{3}{8}$ K. $^{13}$C is rapidly produced by the $\beta^+$ decay
of $^{13}$N with a half life of $9.96$ min, which in turn is the result
of the reaction $^{12}$C(p,$\gamma)^{13}$N that operates here on the
convective turn-over timescale of $\approx 15$ min \citep{herwig:11}.
The He and H burning reactions are spatially separated, each occurring
at its own favourable conditions, and $^{13}$N decays into $^{13}$C
while being carried down by convection, thus avoiding its destruction
by a further proton capture
\citep[e.g.,][]{campbell:10,herwig:11,lugaro:12aipc}.

The first evidence of \iprn\ in stars were observations by \cite{asplund:99} of
$\ga 2$ dex enhancement of the light (first-peak) s-process elements, such as
Rb, Sr and Y, but not of the second-peak elements Ba and La in the post-AGB
very late thermal pulse (VLTP) star Sakurai's object
\citep{Herwig:2001km,Werner:2006bf}. A VLTP is a He-shell flash in a post-AGB
star or young white dwarf (WD) that has already entered the WD cooling track,
i.e.\ the H-burning shell has turned off. He-shell flashes are
commonly and recurrently happening in AGB stars
\citep{Herwig:2005jn}. In the VLTP the He-convection zone reaches into the remaining H-rich
atmosphere, which will mix protons into He-burning conditions
\citep{Fujimoto:1977ww,iben:82}. This mixing triggers \ipr\
nucleosynthesis, and simulations of this situation can reproduce the unusual heavy element \ipr\ fingerprint in Sakurai's
object if mixing predictions from a mixing-length theory (MLT) are modified to
reflect possible effects of the inhomogeneous nature of convective-reactive
burn in H-ingestion events \citep{herwig:11,herwig:14}. The isotopic signature of \iprn\
has since been identified as well in pre-solar grains
\citep{Liu:2014jh,Fujiya:2013ii,Jadhav:2013vc}.

Evidence for \ipr\ nucleosynthesis has also been reported in the anomalous Ba enrichment in
some open clusters \citep{mishenina:15}, while  \cite{2014nic..confE.145D} showed
that nucleosynthesis in \ipr\ conditions provided a remarkable
match to the abundances in several CEMP-r/s (carbon
enhanced metal poor with simultaneous presence of s elements and Eu)
stars.  Possible sites for the \iprn\ include the He-core flash
\citep[e.g.][]{campbell:10}, He-shell flashes in low-Z AGB stars
\citep[e.g.][]{Iwamoto:2004ju} and super-AGB
stars \citep{Jones:2016exa}.

Post-AGB VLTPs from single-star channel are not important sources of heavy
elements on a galactic chemical evolution (GCE) scale. However,
\cite{cassisi:98} reported that $\approx 0.6 M_\odot$ WDs that accreted
solar-composition material at rates of \natlog{4}{-8} and
$10^{-7}M_\odot \mem{/ yr}$ would eventually experience He-shell flashes with
H-ingestion. Their simulations ended due to the ensuing numerical difficulties,
but \cite{cassisi:98} suggested that those He-shell flashes could lead to
substantial additional mass loss as they would cause the WD to evolve back
toward the giant branch, just as the VLTP model of Sakurai's
object. 

This raises several questions. What is the effect of recurrent
He-shell flashes on the mass retention efficiency and the implication
for the single-degenerate supernova of type Ia  \citep[\snia, ][]{hillebrandt:13} 
progenitor channel? What is the
nucleosynthesis signature of rapidly accreting WD (RAWD) He-shell flashes with H
ingestion? Do RAWDs contribute to the chemical inventory of the galaxy.
In this paper we are starting to address these questions through
simulations of RAWDs
for parameters that are similar to those considered for the single-degenerate
progenitor evolution channel of \sneia. Model assumptions and simulations are
presented in \sect{s.assumptions}, followed by our results
(\sect{s.results}) and conclusions (\sect{s.conclusion}).

\section{Model assumptions and simulations}
\label{s.assumptions}

As in our previous work on slowly accreting WDs \citep[the Nova
  Framework,][]{denissenkov:13,denissenkov:14}, we used the
\mesa\ stellar evolution code \citep[here rev.\ 5329,][]{paxton:13} with the same microphysics, except for the reaction
network where we included the pp chains, CNO, NeNa and MgAl cycles, as
well as the main reactions of He burning. The accreted material and
the WD both have the solar initial composition
\citep{grevesse:93,lodders:03}.  Like in our nova models, we have
taken into account convective boundary mixing (CBM) at the bottom of
the He-flash convective zone in some of the evolution tracks via the
exponential, diffusive model
\citep{freytag:96,Herwig:2000ua,Battino:2016bn} implemented in
\mesa. The efficiency parameter $f_\mathrm{bot}$
specifies the e-folding distance of the exponential decay of the mixing efficiency
(\tab{tab:model_summary}).
 
Binary population synthesis (BPS) predicts the formation of close systems,
consisting of a CO WD accretor and a main-sequence (MS) or sub-giant
donor, suitable for the single-degenerate channel to \sneia\ \citep[e.g.,][]{han:04,chen:14}.
For such systems the rate of mass accretion by the WD has to be in a narrow
range around $\dot{M}_\mathrm{acc} \sim 10^{-7} M_\odot \mem{/ yr}$
\citep[e.g.,][]{ma:13,wolf:13}, so that excessive mass loss due to H-shell nova
flashes at lower accretion rates \citep[e.g.,][]{denissenkov:13} or due to
envelope inflation at higher accretion rates is avoided. For our
models we adopt $\dot{M}_\mathrm{acc} \sim 10^{-7} \msol / \mem{yr}$.

Next we have to choose initial values for
the WD and donor (secondary) star mass, $M_\mathrm{WD}$ and $M_\mathrm{d}$,
their orbital separation $a$ or period $P_\mathrm{orb}$, and the WD's
age or  central temperature $T_\mathrm{WD}$. 
According to BPS models the most frequent combinations of the WD and donor star masses at the
beginning of mass transfer in a binary system  from which evolution to \sneia\ may be
 possible are $0.5 \la
M_\mathrm{WD}/M_\odot \la 0.7$ and $0.8 \la M_\mathrm{d}/M_\odot \la
2$, while the suitable orbital periods for the \snia\ outcomes range
from $0.4$ to $6.3\mathrm{d}$, depending on $M_\mathrm{WD}$
\citep{chen:14}. 

We calculate rapid accretion on WD models with $M_\mathrm{WD}\approx
0.65\,M_\odot$ (model A) and $M_\mathrm{WD}\approx 1\,M_\odot$ (model
D) from the Nova Framework \citep{denissenkov:13}. In addition we
construct a new CO WD model with $M_\mathrm{WD}\approx 0.73\,M_\odot$
including the pre-WD evolution starting on the pre-MS (models B
and C).  The WD Roche-lobe radii adopted for our simulations are given
in \tab{tab:model_summary} and correspond to periods of $\sim
0.22\mathrm{d}$ for models A and B and   $\sim
1.4\mathrm{d}$ for model C for a secondary mass $\sim 1\msol$.  

The WD ages when accretion starts follow from basic properties of the
binary evolution scenario, and the corresponding WD central
temperatures of our models are of the order of a few $10^7\mathrm{K}$
(\tab{tab:model_summary}). The WD cools down until its MS or subgiant binary
companion fills its Roche lobe and mass transfer starts. For model C,
an adopted $1.5M_\odot$ MS donor star would expand to its Roche-lobe
radius in 2.426 Gyr. It would take 0.252 Gyr for the $3.7M_\odot$
primary to evolve to the beginning of the WD cooling sequence, where
the WD has the central temperature $T_\mathrm{WD}\approx
10^8\mathrm{K}$. During the remaining $2.174 \mathrm{Gyr}$ between the
two times, the WD would cool down to
$T_\mathrm{WD} < \natlog{1.0}{7}\mathrm{K}$.

Rotation possibly caused by the accretion of angular momentum is not taken into
account. Such rotation may weaken He-shell flashes in a massive and
hot CO WD \citep{yoon:04} and may increase the retention
efficiency of accreted material. We consider less massive and cooler CO WDs
that experience much stronger He-shell flashes, as described
below. The low mass retention efficiency of our models implies that
any accreted angular momentum would be immediately lost again, which
would lessen the effect of rotation.

\section{Results}
\label{s.results}
Models B and C are based on the 
\mesa\ test suite case {\tt make\_co\_wd}. A star with
initially $3.7\,M_\odot$ evolves from the pre-MS
through the usual evolutionary stages to the Asymptotic Giant Branch
(AGB, \fig{fig:fig1}). At that point the star is forced to lose most
of its mass, by artificially imposing a very high mass loss rate, 
as if it went through a common-envelope event. As a result
the star leaves the AGB and evolves into a post-AGB star and then
pre-WD with $M_\mathrm{WD}\approx 0.73 M_\odot$. This model happens to
experience a very late thermal pulse (VLTP, the loop starting at
$\log_{10}T_\mathrm{eff}\approx 5.25$), just like the evolution track for
Sakurai's object. We then let the WD cool down to the adopted core temperature
and start the accretion (\tab{tab:model_summary}) of solar-composition material.

\subsection{The multi-cycle He-flash evolution and the He retention efficiency}
\label{s.multi-cycle}
\begin{figure}
  \centering
  \includegraphics[width=1.\linewidth,clip=True,trim= 0mm 0mm 0mm
  4mm]{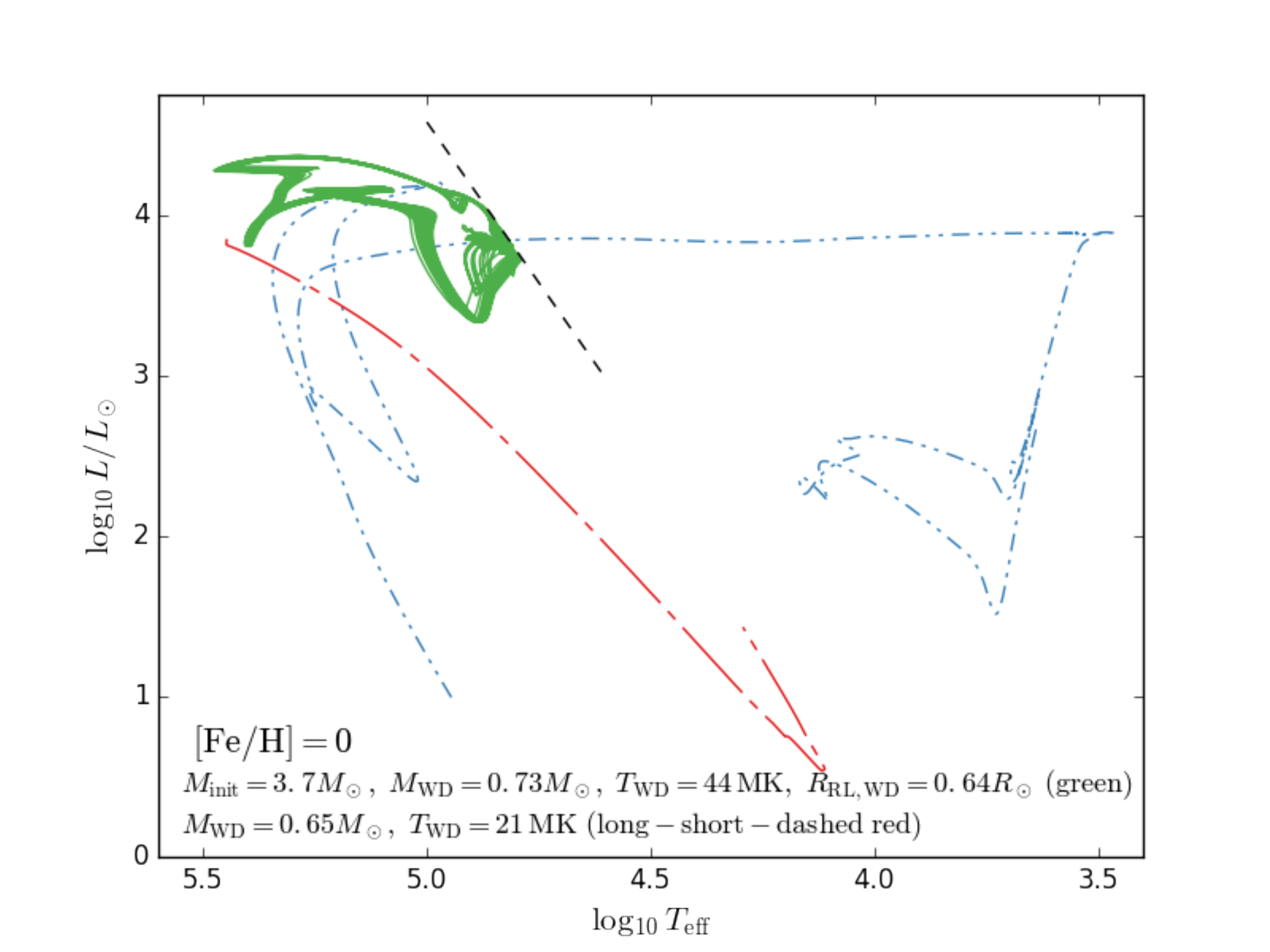}
  \caption{
    Tracks of progenitor and post-AGB evolution of
    $0.73 \msol$ WD (double-dot-short-dashed blue line, see text for details) and
    multiple He-shell flashes with H-ingestion cycles in model B
    (green line). The flash causes the star to expand to the WD
    Roche-lobe radius (dashed black line) and lose the accreted
    material via the Roche-lobe overflow. The short-long-dashed red curve is a fragment of
    the track of model A during its second He-shell flash.}
  \label{fig:fig1}
\end{figure}
When a sufficiently thick layer of He ($\approx 0.012\,M_\odot$ for
models B and C) is accumulated through steady H-shell burning a convective He-shell flash is
triggered (as in model A shown in \fig{fig:fig3}), just as it would be in an AGB stellar
evolution sequence \citep{Herwig:2005jn}. The flash leads to the
expansion of the simulated star just
as in the post-AGB VLTP stellar evolution models
\citep{Werner:2006bf}. Mass loss starts as soon as the stellar model
exceeds the Eddington
luminosity  ($\dot{M}_\mem{wind}
\approx 10^{-4} \msol\mem{/yr}$, model C) or when its radius
reaches the Roche-lobe radius ($\dot{M}_\mem{wind} \approx 10^{-3}
\msol\mem{/yr}$, model B).  Both the accretion and further expansion
pause during the mass loss. When a certain amount of mass is lost,
the model contracts again and the accretion of solar-composition material
resumes. During this duty cycle, the star makes a loop on the HR
diagram (\fig{fig:fig1}). In models C and D the mass-loss is dominated by the super-Eddington luminosity
wind. Yet the peak He-burning luminosities and temperatures remain
quantitatively very similar to those obtained in model B.
\begin{center}
\begin{deluxetable*}{cccccccclr} 
\tablecolumns{8} 
\tablewidth{\columnwidth} 
\tablecaption{
RAWD model parameters \label{tab:model_summary}}
\tablehead{ 
  \colhead{ID}&\colhead{$M_\mathrm{WD}$}&
  \colhead{ $\dot{M}_\mathrm{acc} $}&\colhead{ $T_\mathrm{c} $} &\colhead{ $f_\mathrm{bot} $}
  &\colhead{$R_\mathrm{RL,WD} $}&\colhead{mass} & \colhead{comp. He} & \colhead{$L_\mathrm{He}^\mathrm{max}$} 
&\colhead{ $\eta_\mathrm{He} $}\\
  \colhead{}&\colhead{ $[M_\odot]$}&
  \colhead{ $ [\frac{M_\odot}{\mbox{yr}}]$}&\colhead{ $ [10^7 \mathrm{K}]$}&\colhead{}&
  \colhead{$ [R_\odot]$}&\colhead{loss}&\colhead{flashes}&\colhead{$ [L_\odot]$} &\colhead{\%}}
\startdata 
A & $0.65$ & $\natlog{1}{-7}$ & $21$ & $0$ & $0.65$ & RLOF & $2$ & $\natlog{7.4}{10}$ & $6.8$\\ 
B & $0.73$ & $\natlog{2}{-7}$ & $44$ & $0$ & $0.64$& RLOF & $6$ & $\natlog{4.4}{9}$ & $6.1$ \\ 
C & $0.73$ & $\natlog{2}{-7}$ & $44$ & $0.008$ & $2.2$& $L>L_\mathrm{Edd}$ & $19$ & $\natlog{6.7}{9}$ & $-3.7$ \\
D & $1.00$ & $\natlog{2}{-7}$ & $20$ & $0.004$ & n/a & $L>L_\mathrm{Edd}$ & $4$ & $\natlog{3.2}{10}$ & $10$ 
\enddata 
\end{deluxetable*} 
\end{center}

\begin{figure}
  \centering
  \includegraphics[width=1.\linewidth,clip=True,trim= 0mm 0mm 0mm
  -2mm]{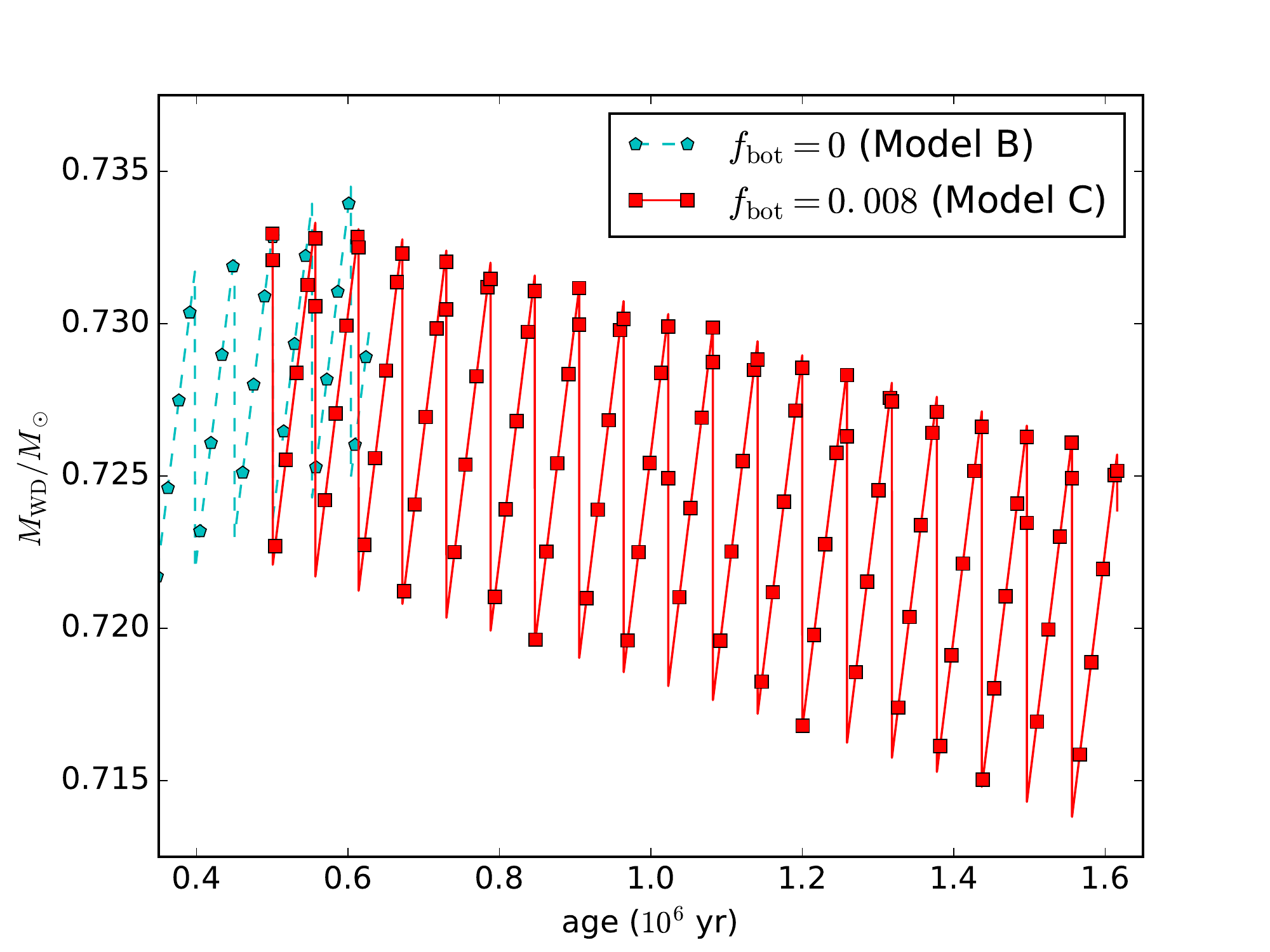}\\
  \caption{The evolution of the total mass of the $0.73 M_\odot$ RAWD
    model B without CBM ($f_\mathrm{bot}=0$) and model C with CBM
    ($f_\mathrm{bot}=0.008$). Each of the saw-tooth shaped features
    represents the mass increase during the accretion phase followed
    by mass ejection in Roche-lobe overflow (model B) or Eddington-luminosity
    mass loss (model C) triggered by the He-shell flash.}
  \label{fig:fig2}
\end{figure}

\begin{figure*}
  \centering
  \includegraphics[width=1.\linewidth,clip=True,trim= 0mm 0mm 0mm
  -2mm]{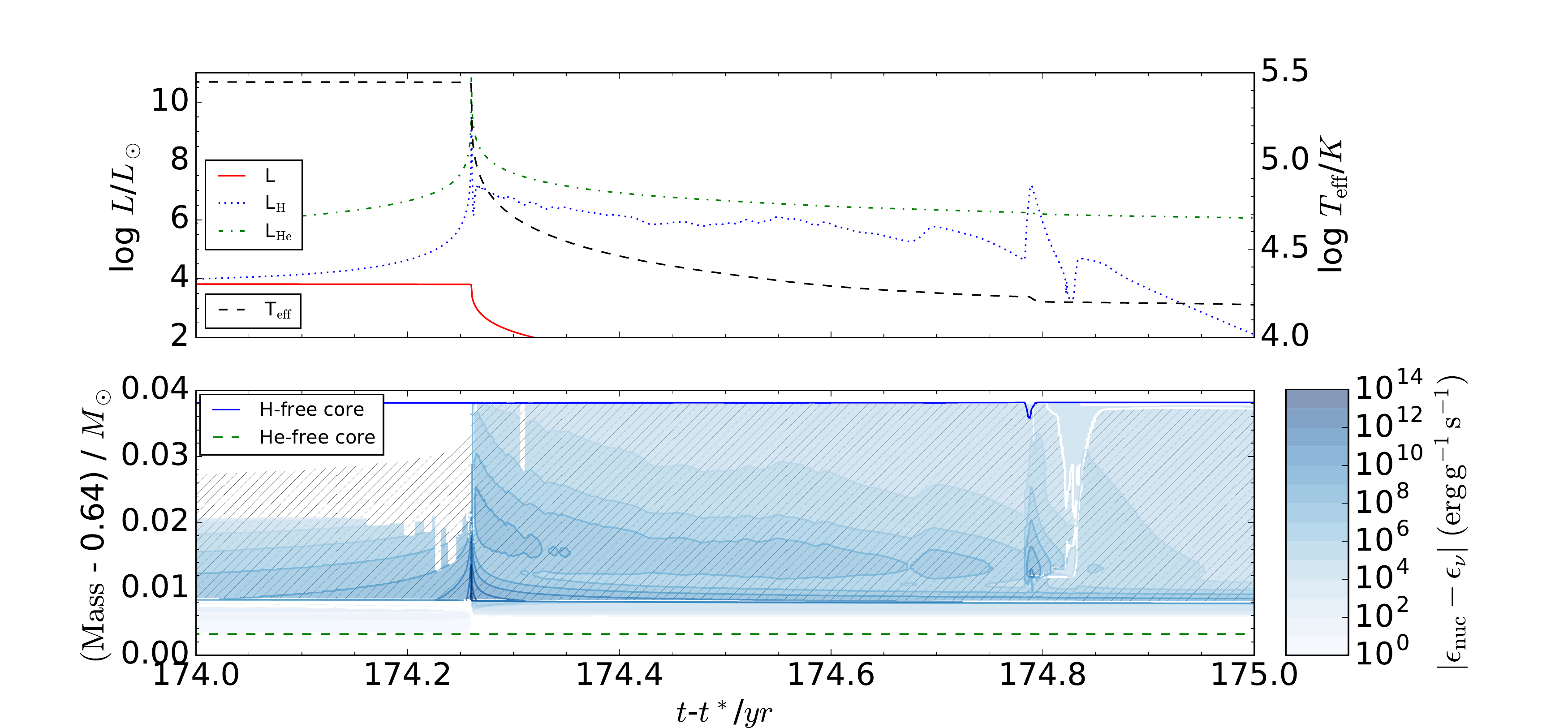}\\
  \caption{Top panel: $T_\mathrm{eff}$ and luminosity evolution
    (total, and due to H and He burning separately). Bottom panel:
    Time evolution of convection zone (hatched area), energy generation (grey shades)
    and H- and He-free mass coordinates diagram of the He-shell flash
    in model A.  $t_{*}$ corresponds to the onset of the flash.  }
  \label{fig:fig3}
\end{figure*}

One of the most important parameters in BPS
studies of the single-degenerate \snia\ progenitor channel is
the He retention efficiency, $\eta_\mathrm{He}$.  It gives a
fraction of the He-shell mass $\Delta M_\mathrm{He}$ that is left on
the WD after the He-shell flash.  It has usually been assumed to be
$100\%$.  This parameter has, to the best of our knowledge, never been
estimated in stellar evolution computations of recurrent He-shell
flashes alternating with phases of steady H burning in rapidly
accreting WDs. The largest He retention efficiency $\eta_\mathrm{He} =
10\%$ is found for the highest mass model (D) with a very small CBM efficiency, while the lower
mass models without CBM (A and B) show $\eta_\mathrm{He} \sim
6\%$ and the $0.73\msol$ model with CBM has $\eta_\mathrm{He} =
-3.7\%$   (Table\,\ref{tab:model_summary} and \fig{fig:fig2}). In the
last case the mass of the WD decreases with time, therefore it cannot
reach the Chandrasekhar mass, unless for some
reason, $\eta_\mathrm{He}$ increases later in the accretion
evolution. Based on numerous investigations of the effect that CBM at the
bottom of the He-shell flash convection zone
has on stellar evolution simulations
\citep[e.g.][]{Herwig:2005jn,Werner:2006bf,MillerBertolami:2006dr,Weiss:2009khb,Karakas:2010dk,Battino:2016bn},
its investigation in 3D simulations \citep{Herwig:2007wh} and the
better agreement of observations of novae with models that include CBM
at the bottom of the H-flash convection \citep{denissenkov:13}, we
conclude that CBM with $f_\mathrm{bot}=0.008$ at the bottom of the
He-shell flash convection zone is at present the most realistic
assumption. 

\subsection{The \iprn}

The He-shell flashes that we encounter in the rapidly accreting white dwarf (RAWD) models are similar
to the post-AGB He-shell flash model that matches the observations of
Sakurai's object, in the sense that also in the RAWD models H from the
envelope ($\Delta M_\mathrm{env}\sim 10^{-5}$ to $10^{-4}\,M_\odot$)
is ingested into the He-shell flash convection (\fig{fig:fig3}).  The
H-ingestion events are seen in all of our simulated He-shell flashes.

The ingested protons react with \czw\ to form
\ndr. Over its $\beta^+$-decay time scale it is advected toward
the bottom of the He-shell flash convection zone, where
temperatures are high enough to activate a rapid release of neutrons
via the $\cdr(\alpha,\n)\ose$ reaction. Just as in the case of
Sakurai's object \citep{herwig:11}, H-ingestion events in RAWD models lead to
\ipr\ nucleosynthetic abundance patterns. Given that most or
all of the material from the He-shell
convective zone is ejected (\sect{s.multi-cycle}), RAWDs may be important sources of
enrichment of the ISM in heavy elements produced in the i process
(\sect{s.conclusion}).

One important difference between the H-ingestion RAWD models and the VLTP or
low-Z AGB models is a result of the WD age and associated compactness. This
causes He-shell flashes in RAWDs to be very strong. The peak He-burning
luminosity is $L_\mathrm{He,peak}= 10.869$ and remains for most of the
H-ingestion event much higher than the energy released by H burning which
reaches $L_\mathrm{H,peak}= 9.673$ (\fig{fig:fig3}). In spite of this enormous
energy input into the He-shell flash convection zone from the burning of
ingested H, the stellar evolution simulation does not indicate the formation of
a split of the convection zone and the formation of a separate H-burning driven
convection as has been commonly observed in H-ingestion simulations of post-AGB
and low-Z AGB stars (\sect{s.introduction}).

During the initial phase of the
He-shell flash of model A at $t=174.25\mathrm{yr}$, we first observe a very high
H-ingestion rate followed by a prolonged period of lower H ingestion,
and all of this without split of the He-flash convection zone. If it
is confirmed in 3D simulations that no Global Oscillation of Shell
H-ingestion \citep[GOSH,][]{herwig:14}  is found in this
situation, 1D spherically symmetric simulations of the
\ipr\ nucleosynthesis in RAWDs are probably more realistic compared to
the case of post-AGB VLTP models and low-Z AGB models for which 3D
stellar hydrodynamics simulations have yet to determine the detailed
outcome of the GOSH.

At a later time ($t=174.8\mathrm{yr}$) the
H-ingestion rate does increase for a brief period when the convection
zone is indeed split. Whether or not this particular event is due to
the decreasing He-burning luminosity, and how the detailed properties
of this event depend on numerical and physics model assumptions is not
yet clear. However, our simulations suggest that there is a distinct
possibility of diversity in the convective mixing and nuclear burning
interactions in H-ingestion events on degenerate cores. 

We adopt the same 1D multi-zone post-processing nucleosynthesis simulation approach as in \citet[][for details
see \S\,5.1 there]{herwig:11} using the NuGrid \mppnp\ code \citep{Pignatari:2016er}.  The initial abundances for the simulations are taken
from \cite{asplund:05} with isotopic ratios from \cite{lodders:03}.  The
initial abundances of some light intershell species, such as \hefo, \czw, \ose\
are modified to reflect their intershell abundance due to the progenitor AGB
evolution. The \mppnp\ code dynamically includes up to $5234$ isotopes as
needed and the associated reaction rates from JINA \textsc{REACLIB v1.1}
\citep{cyburt:10} and select other sources \citep{Pignatari:2016er}.

We perform simulations for an initial prolonged ingestion ($\sim 0.38
\mathrm{yr}$) without split with a low ingestion rate ($\sim
\natlog{1.8}{-12} \msol \mem{/s}$) as well as the subsequent shorter
event ($\sim 0.0084 \mathrm{yr}$) with a high ingestion rate ($\sim
\natlog{8.9}{-11}\msol / \mem{s}$) that does lead to a
convection zone split. The $T$-$\rho$ profile of the He-shell
convective zone was taken from model A 
when the H-ingestion has just begun during
the second He-shell flash. 

\begin{figure}
  \centering
  \includegraphics[width=1.\linewidth,clip=True,trim= 0mm 0mm 0mm
    -2mm]{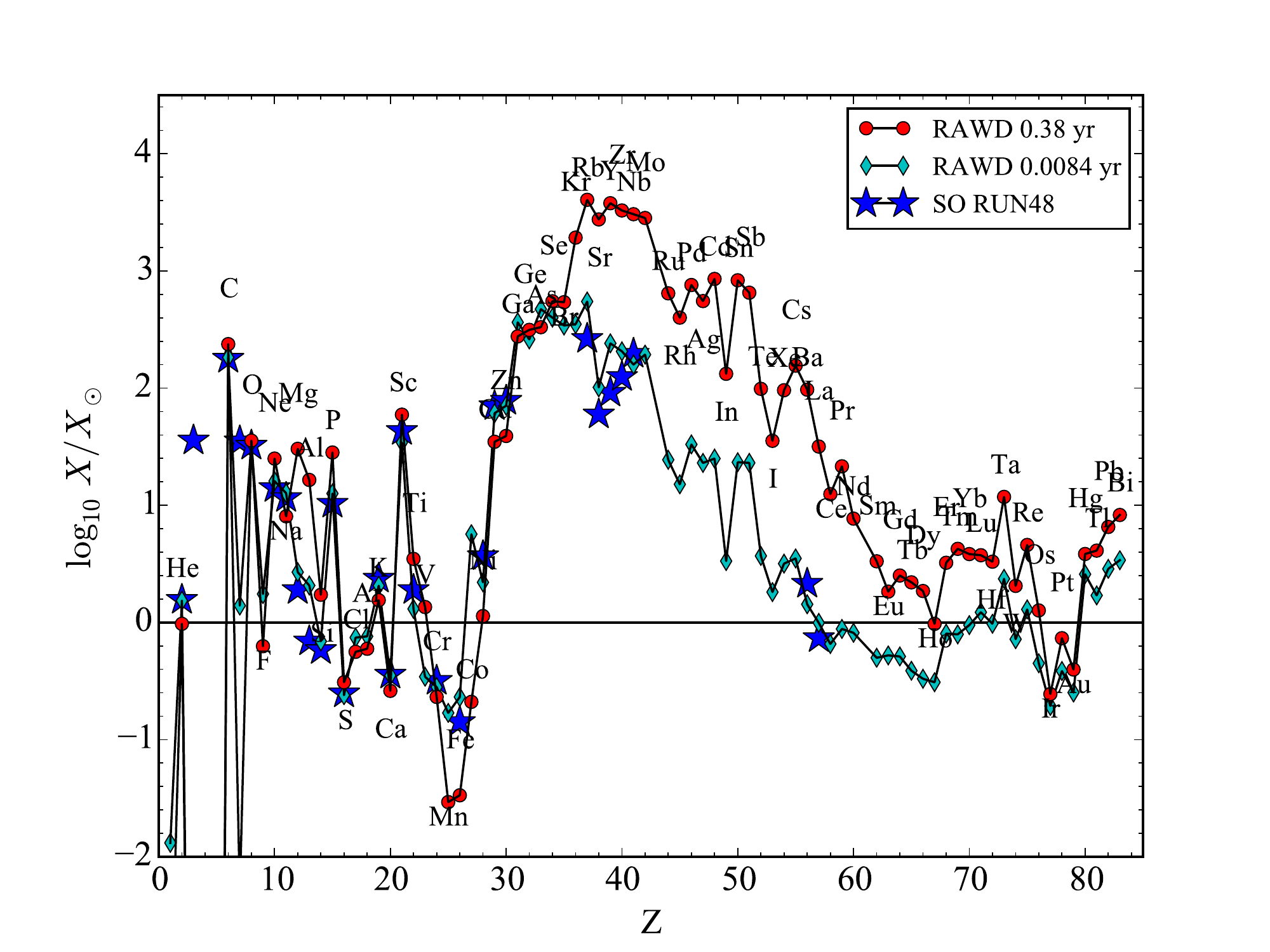}
  \caption{
Abundance distributions in the RAWD model A, after the second He flash, with
    $M_\mathrm{WD}=0.65\,M_\odot$ and $T_\mathrm{WD}=21 \mathrm{MK}$
    (teal diamonds, solid lines),
    and for comparison the model RUN48 (blue stars) from
    \citet{herwig:11} that matches the observed abundances of
    Sakurai's object.  }
  \label{fig:fig4}
\end{figure}

In both simulations a substantial overproduction of mostly first-peak
trans-iron elements is found (\fig{fig:fig4}). The abundance signature
of the short event (H-ingestion with split) is very similar to the
post-AGB He-shell flash simulation that matches the observed abundance distribution in
Sakurai's object. The longer event (without split) shows even higher
overproduction up to 3.5 dex and the abundance distribution is
starting to spill over the first peak $N=50$ magic neutron number,
generating elements such as Mo and Ag. These simulations demonstrate
that the ejecta of RAWDs may be enriched with first-peak heavy
elements by 2.5 to 3.5 dex, and that a range of conditions may lead to
a certain diversity of local elemental ratios within the \ipr\ paradigm.

\section{Conclusion}
\label{s.conclusion}

\cite{iben:81} proposed that accreting WDs may be a
potential astrophysical site for the s\,process, via 
the $^{22}$Ne($\alpha$,n)$^{25}$Mg reaction in He-shell flashes. It was later realized
that the main s process is best associated with the radiative
\cdr\ neutron source \citep{Gallino:1998eg} in AGB stars. Our models show that for WD
accretion models with $M_\mathrm{WD}\sim 0.7-0.8 M_{\odot}$ the
nucleosynthesis signature should instead be dominated by the
\iprn\ fueled by the convective $\cdr(\alpha,\n)\ose$ neutron source.

The Milky-Way present-day star-formation rate is
$\sim 2\,M_\odot/\mem{yr}$
\citep{chen:14}. A fraction of $\sim
  0.08$ of this will go into low-mass AGB stars
  ($1.5\msol \leq M_\mathrm{ini} \leq 3\msol$) that produce the main s process, and they
  return $\sim 70\%$ of their mass to the ISM. Therefore,
  $\approx
0.1\msol \mem{/yr}$ of s-process enriched material is returned by
low-mass AGB stars. This material is enriched by $\sim 2$ compared to the initial abundance of heavy elements
\citep[e.g.][]{lugaro:03}.

As a lower limit, we adopt for the RAWD rate the presently estimated rate of \sneia\ from
the single-degenerate channel, $2\times 10^{-4}\,\mathrm{yr}^{-1}$ \citep{chen:14}. If one assumes
that for each RAWD $\sim 0.5\,M_\odot$ of H-rich material can be
accreted, and ejected enriched with i-process elements (assuming
$\eta_\mathrm{He} =0$) then $\approx 0.0001\msol/\mathrm{yr}$ of
i-process enriched material is returned by RAWDs with an enrichement
factor $\sim
1000$ (\fig{fig:fig4}).

The ratio of the contributions of elements made by both low-mass AGB stars and RAWDs is then
$$
\frac{\mathrm{AGB\ contribution}}{\mathrm{RAWD\ contribution}} =  \frac{0.1\msol/\mathrm{yr}}{0.0001\msol/\mathrm{yr}}\times \frac{2}{1000} = 2.0 \mathrm{\, .}
$$
Therefore, based on results shown in \fig{fig:fig4}, we propose that the \iprn\
in RAWDs may be a relevant astrophysics source for elements in
the Ge-Mo region. The i process in RAWDs could be one of the
nucleosynthesis components explaining the missing solar LEPP
 \citep{travaglio:04,Montes:2007dm}.
 
Another important result is that,  according to our stellar evolution
models of RAWDs that follow through their He-shell flashes, the He
retention efficiencies are $\la
10\%$, or negative in models with convective boundary mixing. This result is
consistent with the recent He-accreting WD models \citep[][and references therein]{wang:15}.
BPS estimates of the \snia\ rate from the single-degenerate channel usually assume a
He retention efficiency of $100\%$ \citep[e.g.,][]{chen:14}. This means
that according to our models this channel for \sneia\ via RAWDs is
highly unlikely, except possibly for a very small fraction of systems
with the most massive WDs and donor stars. RAWD systems  are predicted by numerous population synthesis models.  They should appear as super-soft X-ray sources for the most time,
unless being (easily) obscured by interstellar or circum-binary matter \citep{vandenheuvel:92}. 
The latter factor probably explains why only a few RAWD candidates out of theoretically predicted dozens
were found in the LMC and SMC \citep{lepo:13}.

\acknowledgments 
This material is based upon work supported by the
National Science Foundation under Grant No. PHY-1430152 (JINA Center for the
Evolution of the Elements). 
FH acknowledges funding from an NSERC Discovery grant. 
MP acknowledges the support from SNF (Switzerland) and 
from the "Lendulet-2014" Programme of the Hungarian Academy of Sciences (Hungary).
SJ is a fellow of the Alexander von Humboldt Foundation and acknowledges support from
the Klaus Tschira Foundation.


\end{document}